\documentclass{elsart}

\usepackage{epsfig}

\usepackage{amssymb}

\def\Hz{\rm\, Hz}

\def\bfl{\mathbf{\pi}\rm }

\def\bfe{\mathbf e\rm}
\def\bfx{\mathbf  x\rm}
\def\bfp{\mathbf  p\rm}

\def\bfa{\mathbf  a\rm}

\def\bfd{\mathbf  d\rm}

\def\bfD{\mathbf  D\rm}
\def\bfb{\mathbf  b\rm}

\def\bfX{\mathbf  X\rm}
\def\bfP{\mathbf  P\rm}
\begin{document}

\begin{frontmatter}



\title{A modified Least Squares Lattice filter\\ to identify non stationary process}


\author{Elena Cuoco}
\ead{cuoco@fi.infn.it}

\address{Istituto Nazionale di Fisica Nucleare Firenze,\\ 
Via G. Sansone 1, 50019 Sesto F. Firenze, Italy}

\begin{abstract}
In this paper the author proposes to use the Least Squares Lattice filter with forgetting factor to estimate time-varying parameters of the model for noise processes. We simulated an Auto-Regressive (AR) noise process in which we let the parameters of the AR vary in time. 
We investigate a new way of implementation of Least Squares Lattice
filter in following the non stationary time series  for stochastic process.
Moreover we introduce a modified Least Squares Lattice filter to whiten the time-series and to remove the non stationarity. 
We apply this algorithm to the identification of real times series data produced by recorded voice.
\end{abstract}

\end{frontmatter}

\section{Introduction}
\label{sec:intro}
In the application of optimal filtering for the detection of a signal buried in noise, often it is useful the procedure of whitening~\cite{cuoco1,cuoco2}. If the noise in which the signal is hidden is non stationary or non Gaussian noise we cannot apply anymore the optimal filter for stationary and Gaussian noise.
We focused this work on the application of Least Squares Lattice~\cite{Haykin,Alex,Bouchard,Kim} algorithm to the problem of identification of non stationary noise in stochastic process, to the procedure of whitening and to the possibility of making stationary a non stationary process.

To this aim we apply this algorithm to some toy models we built using an autoregressive non stationary model (see section~\ref{sec:toy})~\cite{Kay,Zheng,Dickie}. In
section~\ref{sec:whitening} we show the whitening techniques based on a
lattice structure, in section~\ref{sec:lsl} we introduce the adaptive  Least Squares methods and its application to simulated non stationary data. In sections~\ref{sec:toy} and~\ref{sec:voice}  we show how it is possible to whiten the data and to eliminate the non stationarity present in the noise. We apply this algorithm to simulated and real data.

\section{The autoregressive model and the whitening}
\label{sec:whitening}
An Auto-Regressive process $x[n]$ of order $P$ with parameter $a_k$, from here after $AR(P)$, is characterized by the relation
\begin{equation}
\label{eq:AR}
x[n]=\sum _{k=1}^{P}a_{k}x[n-k]+\sigma w[n]\, ,
\end{equation}
being $w[n]$ a white Normal process.

The problem of determining the AR parameters is the same of that of finding
the optimal ``weights vector'' ${\mathbf w}=w_k$, for $k=1,...P$  for the
problem of linear prediction~\cite{Kay}. 
In the linear prediction we would predict the sample $ x[n] $ using the $ P $
previous observed data ${\mathbf x}[n]=\{x[n-1],x[n-2]\ldots x[n-P]\} $
building the estimate as a transversal filter:
\begin{equation}
\hat{x}[n]=\sum _{k=1}^{P}w_{k}x[n-k]\, .
\end{equation}
 
We choose the coefficients of the linear predictor by minimizing  a cost
function that is the mean squares
error $ \epsilon ={\mathcal{E}}[e[n]^{2}] $ (${\mathcal{E}}$ is the operator of average on the ensemble), being 
\begin{equation}
\label{eq:error}
e[n]=x[n]-\hat{x}[n]
\end{equation}
the error we make in this prediction and obtaining the so called Normal or
Wiener-Hopf  equations
\begin{equation}
\label{eq:Normal}
\epsilon _{min}=r_{xx}[0]-\sum _{k=1}^{P}w_{k}r_{xx}[-k]\, ,
\end{equation}
 which are identical to the Yule--Walker equations~\cite{Kay} used to estimated the parameters $a_k$ from autocorrelation function with $w_{k}  = -a_{k}$ and $\epsilon _{min}  =  \sigma ^{2}$

This relationship between AR model and linear prediction assures us to obtain
a filter which is stable and causal~\cite{Kay}, so we can use the $AR$ model to reproduce stable processes in time-domain. It is this relation between
AR process and linear predictor that becomes important in the building of whitening
filter.

\begin{figure}
\begin{center}
\epsfig{file=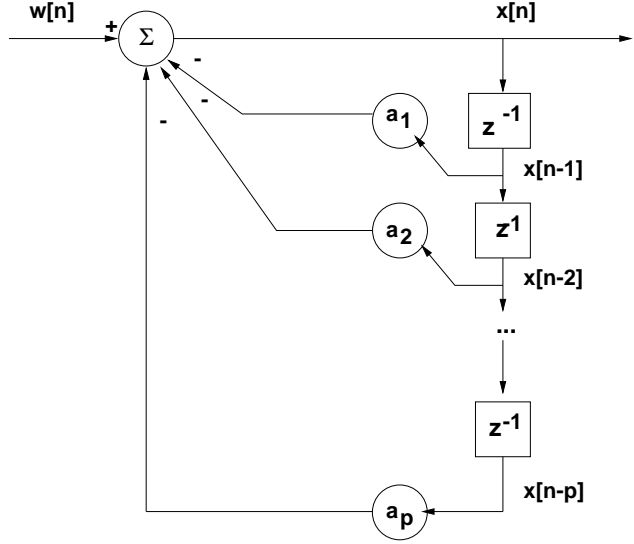,width=0.6\textwidth}
\end{center}
\caption{Whitening filter and AR filter.}
\label{fig:ar-predic}
\end{figure}

The tight relation between the AR filter and the whitening filter is clear in
the figure~\ref{fig:ar-predic}. The figure describes how an AR process colors
a white process at the input of the filter if you look at the picture from left
to right. If you read the picture from right to left you see a colored process
at the input that pass through the AR inverse filter coming out as a white process.

When we find the $ P $ parameters that fit a PSD of a noise process, what
we are doing is to find the optimal vector of weights that let us reproduce the process
at the time $ n $ knowing the process at the $ P $ previous times. All
the methods that involve this estimation try to make the error signal (see
equation (\ref{eq:error}) ) a white process in such a way to throw out all the correlation
between the data (which we use for the estimation of the parameters).

\section{LSL filter}
\label{sec:lsl}

\begin{figure}
\begin{center}
\epsfig{file=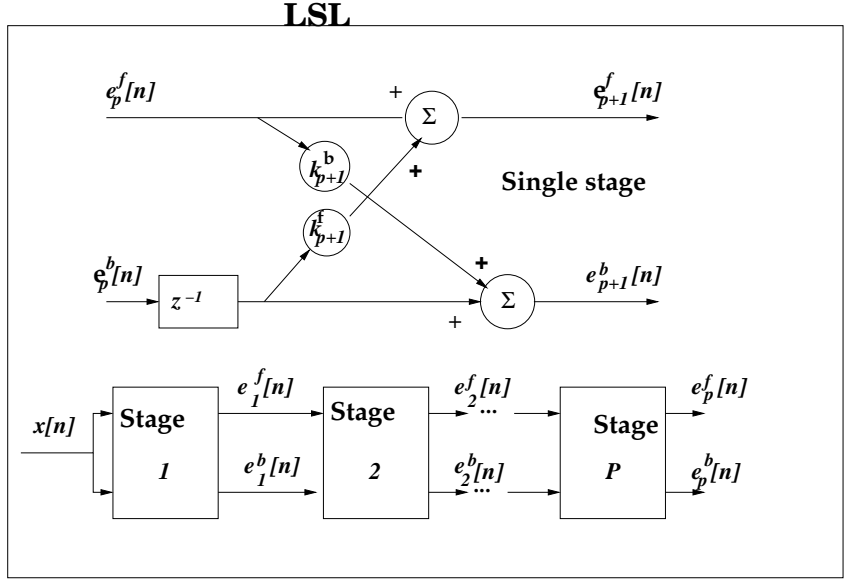,angle=0,width=0.8\textwidth}
\end{center}
\caption{Lattice structure for LS filter.}
\label{fig:lattice}
\end{figure}

The Least Squares based methods build their cost function using all the information
contained in the error function at each step, writing it as the sum
of the error at each step up to the iteration $n$ : 
\begin{equation}
\label{rls1}
\epsilon [n]=\sum _{i=1}^{n}\lambda ^{n-i}e^{2}(i|n)\, ,
\end{equation}
 
being 
\begin{equation}
\label{rls2}
e(i|n)=d[i]-\sum _{k=1}^{N}x_{i-k}w_{k}[n],
\end{equation}

where $ d $ is the  signal to be estimated, $ x $ are the data
of the process and $ w $ the weights of the filter.
The forgetting factor $ \lambda  $  lets us tune the
learning rate of the algorithm. This coefficient can help when there are non
stationary data in the time series and we want the algorithm has a short memory.
If we have stationary data we fix $ \lambda =1 $, otherwise we choose $0<\lambda <1$ 

There are two ways to implement the Least Squares methods
for the spectral estimation: in a recursive way (Recursive Least Squares or
Kalman Filters) or in a Lattice Filters using fast techniques~\cite{Alex}.
 The first kind of algorithm, examined in~\cite{cuoco1}, has a computational cost
proportional to the square of the order of filter, while the cost of the
second one is linear in the order $ P. $

The computational cost of RLS is prohibitive for an on line implementation.
Moreover its structure is not modular, thus forcing
the choice of the order $P$ once for all.
The algorithm with a modular structure like that of the lattice  offers the
advantages of giving an output of the filter at each stage $p$,  so  in
principle we can change the order of the filter by imposing some criteria on
its output. The Least Square Lattice filter is a modular
filter with a computational cost proportional to the order $P$.

In the least squares methods the linear prediction is made for a vector
of data $\hat{\bfx}[n]$, so the natural space where developing these methods are the vectorial spaces (a detailed insight in these techniques is reported in~\cite{Alex}).

Let ${\bf X}$ be a Hilbert $p-$dimensional space, to which  the vectors  
$\bfx[n]$ of acquired data belong. The $p$ vectors   $\bfx_j[n]$ of length
$n$ obtained as time translation of length $p$ of the vector 
$\bfx[n]$
\begin{eqnarray}
  \label{eq:shift}
  \bfx_1[n]=z^{-1}\bfx[n]&=&(0,0,x[1],...,x[n-1])\ ,\nonumber\\
  \bfx_2[n]=z^{-2}\bfx[n]&=&(0,0,0,x[1],...,x[n-2])\ ,\nonumber\\
  \vdots &=& \vdots \nonumber\\
  \bfx_p[n]=z^{-p}\bfx[n]&=&(0,0,0,\ldots,x[1],...,x[n-p])\nonumber
\end{eqnarray}
form a base of this space.
A vector $\bf u$ which belongs to this space can be written as
\begin{equation}
{\bf u}[n]=\sum_{k=1}^{p}a_k\bfx_k[n]\, .
\end{equation}   
The vector $\bfx_{p+1}[n]$ with last component $x[n]$ does not belong to this
space, but to a vectorial  $p+1$-dimensional space ${\bf D}$.
In the problem of linear prediction the best estimation of desired signal
$\bfd[n]$, that is $\bfx_{p+1}[n]$, is obtained using the vector lying in the
space  ${\bf X}$.

Therefore the  least squares methods look for a vector
 $\hat{\bfx}[n]$ which is the closest to the vector $\bfd[n]$, by minimizing the norm of the
distance between $\hat{\bfx}[n]$ and $\bfd[n]$.  It can be shown that this
operation corresponds to the projection of the vector $\bfd[n]$
from the  $p+1$-dimensional space $\bfD$  in the  $p$-dimensional sub-space
$\bfX$   by a projector ${\bf P}$.
We can decompose the vector $\bfd[n]$ as sum of the vector
$\hat{\bfx}[n]$ and a vector which has null component only along the vector
orthogonal to the space $\bfX$.  This vector is the vector
${\bf e}(n|n)$ which, by definition, is orthogonal to the data vector
$\bfx[n]$. In fact the orthogonal vector to the space $\bfX$ can be obtained as

\begin{equation}
  \label{eq:eproiet}
  ({\bf I}-{\bf P})\bfd[n]=\bfd[n]-{\bf P}\bfd[n]=\bfd[n]-\hat{\bfx}[n]={\bf e}(n|n)\ .
\end{equation}

Therefore the vector $\bfd[n]$ belongs to the vectorial space 
${\bf D}$,  direct sum of the sub-space ${\bf X}$ and of sub-space ${\bf E}$  defined by
the vector ${\bf e}(n|n)$

\begin{equation}
  \label{eq:vettoriali}
  {\bf D}=\bfX\oplus{\bf E}\, .
\end{equation}
For the LS adaptive algorithm we want to write the quantities we need for the
estimation of $\hat{\bfx}[n]$ at the iteration $n$ by means of the quantities
at the iteration $n-1$ and if we use a modular structure the same quantities
at the stage $p$ in terms of the ones at the stage $p-1$.

Using the described techniques, if  we augment the order of the filter 
from  $p$ to $p+1$, we must write the new projector ${\bf P}_{p+1}$ as function
of the operator ${\bf P}_{p}$.
The new vectorial space will be the direct sum of 
${\bf X}$ of dimension  $p$ and of the  $1$-dimensional sub-space   orthogonal
to ${\bf X}$ along which there is ${\bf e}(n|n)$ and the projector will be 
\begin{equation}
  \label{eq:Proiec_new}
 {\bf P}_{p+1}={\bf P}_p + {\bf P}_1\ ,
\end{equation}
where we wrote $\bfP_1$ to point the projector on the  one-dimensional
space $\perp \bfX$.

If we add a new data  $x[n]$ to the space  ${\bf X}(n-1)$,
we introduce the vector $\bfl[n]$ orthogonal to the space ${\bf
  X}(n-1)$ and the new projection of the signal $\bfd[n]$ along  ${\bf X}(n)$ 
will be

\begin{eqnarray}
  \label{eq:proiec_time}
  {\bf P}[n]\bfd[n]=\bfP[n-1]\bfd[n-1]+{\bf P}_{\bfl[n]}\bfd[n]=\\ \nonumber
\hat{\bfx}[n-1] +{\bf P}_{\bfl[n]}\bfd[n]\ .
\end{eqnarray}
Then we can write in a matrix form the relation~(\ref{eq:proiec_time})
\begin{equation}
  \label{eq:Proiectn-n-1}
  {\bf P}[n]=\left(\begin{array}{ll}
      {\bf P}[n-1]&0\\
      0&1
\end{array}\right)\, .
\end{equation}

A useful parameter which is introduced is the angle $\gamma_p[n-1]$ between the two sub-spaces  ${\bf X}[n-1]$ and 
${\bf X}[n]$ which can be obtained from the relation
\begin{equation}
  \label{eq:gamma}
  \gamma_p[n-1]=<\bfl[n],{\bf P}^\perp_p[n]\bfl[n]>\ ,
\end{equation}
where we introduced  the scalar product $<,>$ between the two vectors
$\bfa[n]$ $\bfb[n]$ defined as
\begin{equation}
  \label{eq:scalare}
  <\bfa[n],\bfb[n]>=\sum_{k=1}^{n}\lambda^{n-k}u[k]v[k]\ ,
\end{equation} 
and ${\bf P}^\perp_p[n]={\bf I}-{\bf P}_p[n]$. Let us remember that 
$\lambda$ is the   forgetting factor; if we limit ourselves to 
$\lambda=1$ the scalar product  $<,>$ is simply
$\bfa^T\bfb$.  

We can write an adapting relation for the projector in term of the vector  $\gamma$ 
\begin{equation}
  \label{eq:adattPperp}
  \left[\begin{array}{ll}
\bfP_p^\perp[n-1]& 0\\
0&0
\end{array}\right]=\bfP_p^\perp[n]-\frac{\bfP_p^\perp[n]\bfP_{\bfl}[n]
\bfp_p^\perp[n]}{\gamma_p[n]}\ .
\end{equation}

It is important to note that, thanks to the properties of the projector, 
the number of operation per iteration is now proportional to the order $P$ and not
to $P^2$ as for the RLS algorithm.

The  LSL filter is a lattice filter  characterized by recursive relation between the forward error (FPE) and
the backward one (BPE). With the new notation we can write

\begin{equation}
  \label{eq:FPElsl}
  \bfe_p^f[n]=\bfx[n]-\hat{\bfx}[n]=\left[{\bf I}-{\bfP}_p[n]\right]\bfx[n]\ .
\end{equation}

The scalar error $e_p^f[n]$ can be written as the component along the
direction $\bfl[n]$ of the vector perpendicular to the sub-space $\bfX[n-1]$
\begin{equation}
  \label{eq:FPE_per}
  e^f_p=<\bfl[n],\bfe_p^f[n]>\ .
\end{equation}
In a similar way we can write the backward errors. For the backward errors the
space where we make the prediction is different from the sub-space $\bfX[n]$ 
because the base is now given by
$z^0\bfx[n],\ldots,z^{-(p-1)}\bfx[n]$.  If we introduce the projector
$\bfP_{p-1}$ on this new base we can write
 \begin{equation}
  \label{eq:BPElsl}
  \bfe^b_p[n]=\left[ {\bf I}-\bfP_{p-1}[n]\right]z^{-p}\bfx[n]\ .
\end{equation} 
The scalar backward error is given by
\begin{equation}
  \label{eq:FBE_scalar}
  e_p^b[n]=<\bfl[n],\bfe_p^b[n]>\, .
\end{equation}
The square sum for the forward and backward errors can be written as 
\begin{eqnarray}
  \label{eq:FPEBPE}
  \epsilon_p^f[n]&=&<\bfe_p^f[n],\bfe_p^f[n]>\\
  \epsilon_p^b[n]&=&<\bfe_p^b[n],\bfe_p^b[n]>
\end{eqnarray}
Now we can write the recursive relations for the projectors in the equations 
~(\ref{eq:FPElsl}) and~(\ref{eq:BPElsl}) 
\begin{eqnarray}
  \label{eq:reticolo}
e^f_{p+1}[n]&=&e^f_p[n]+k_{p+1}^b[n]e^b_p[n-1]\ ,\\
 \label{eq:reticolo1}
e^b_{p+1}[n]&=&e^b_p[n-1]+k_{p+1}^f[n]e^f_p[n]\ ,
\end{eqnarray}
where we introduced the  forward $k^f_p$ and
 backward $k^b_p$ reflection coefficients defined by
\begin{eqnarray}
  \label{eq:kpfkpb}
  k^b_{p+1}[n]&=&-\frac{<z^{-1}\bfe_p^b[n],\bfe_p^f[n]>}{\epsilon^b_p[n-1]}\ ,\\
& &\nonumber\\
 k^f_{p+1}[n]&=&-\frac{<z^{-1}\bfe_p^b[n],\bfe_p^f[n]>}{\epsilon^f_p[n]}\ .
\end{eqnarray}
The adaptive implementation of the LSL
filter~(\ref{eq:reticolo})~(\ref{eq:reticolo1}) requires the updating with
order $p$ and time  $n$  of the reflection coefficients.

So we must write the recursive relation for the quantities
$\epsilon_p^f[n]$, $\epsilon_p^b[n]$ and $\Delta_{p+1}[n]=<z^{-1}\bfe_p^b[n],\bfe_p^f[n]>$.  
This can be done using the updating formula
\begin{eqnarray}
\Delta_{p+1}[n]&=&\lambda\Delta_{p+1}[n-1]+
\frac{e^b_{p}[n-1]e^f_{p}[n]}{\gamma_p[n-1]}\ ,\\
\epsilon^f_{p+1}[n]&=&\lambda\epsilon^f_{p}[n]-
\frac{\Delta^2_{p+1}[n]}{\epsilon_p^b[n-1]}\ ,\\
\epsilon^b_{p+1}[n]&=&\lambda\epsilon^b_p[n-1]-
\frac{\Delta^2_{p+1}[n]}{\epsilon_p^f[n] }\ , \\
\gamma_{p+1}[n-1]&=&\gamma_p[n-1]-
\frac{[e_p^b[n-1]]^2}{\epsilon_p^b[n-1]}\ .
\end{eqnarray}

The error $e_p^f$ at the last stage is the whitened sequence of the input data.
So at the output of LSL filter we find the reflection coefficients we can use for the estimation of the
AR parameters for fit to  PSD of the time series. Moreover one of the output of this filter is  the whitened sequence of data.

The procedure described for the implementation of LSL filter is the so called aposteriori procedure~\cite{Haykin}. Since this algorithm involves division by updated parameters, we must be careful in avoiding division by values too small. In the aposteriori procedure, the reflection coefficients are estimated indirectly by the estimation of $\epsilon^f$ and  $\epsilon^b$.

In the apriori implementation, the reflection coefficients are estimated directly by the forward and backward errors. 

In the apriori implementation the recursive relation for the parameters are given 
by
\begin{eqnarray}
\epsilon^f_{p-1}[n] & = & \lambda \epsilon^f_{p-1}[n-1]+ \gamma_{p-1} e^f_{p-1}[n] e^f_{p-1}[n]\ ,\\
\epsilon^b_{p-1}[n] & = & \lambda \epsilon^b_{p-1}[n-1]+ \gamma_{p-1} e^b_{p-1}[n-1] e^b_{p-1}[n-1]\ ,\\
e^f_{p} [n]&=& e^f_{p-1}[n]+ k^f_{p}[n-1] e^b_{p-1}[n-1]\ , \\
e^b_{p}[n] &=& e^b_{p-1}[n-1]+  k^b_{p}[n-1] e^f_{p-1}[n]\ , \\      
k^f_{p}[n] &=& k^f_p[n-1]-\frac{\gamma_{p-1}e^b_{p-1}[n]e^f_{p}[n]}{\epsilon^b_{p-1}[n]}\ , \\
k^b_p[l] &=& k^b_p[n-1]-\frac{\gamma_{p-1}e^f_{p-1}[n]e^b_{p}[n]}{\epsilon^f_{p-1}[n]}\ , \\
\gamma_{p}&=&\gamma_{p-1}-\frac{\gamma_{p-1}^2|e^b_{p-1}[n-1]|^2}{\epsilon^b_{p-1}[n]}\ . \\
\end{eqnarray}
Since the apriori implementation is more stable with respect the aposteriori one, we choose to use the apriori recursive relation for the LSL filter to perform tests on non stationary data.

\section{ Modification of LSL filter output to remove non stationarity in the data}

\begin{table}
  \caption{Modified LSL algorithm}
  \label{tab:lsl}
   \begin{tabular}{l}
\hline
Parameter and variable descriptions \\
 $l$: time index \\
 $p$: index on filter stage \\
$ x[l]$: input data sequence \\
 $y[l]$: whitened data sequence \\
Main Loop \\
for $l=1,2,...N$\\
$e^b_0[l]=e^f_0[l]=x[l]$\\
$\epsilon^f_0[l] = \epsilon^b_0[l]=\lambda \epsilon^f_0[0]+x^2[l] $ \\
$\gamma[0]=1.0$\\

for $p=1,2, P$  \\
$ \epsilon^f_{p-1}[n]  =  \lambda \epsilon^f_{p-1}[n-1]+ \gamma_{p-1} e^f_{p-1}[n] e^f_{p-1}[n]\, $\\
$\epsilon^b_{p-1}[n]  =  \lambda \epsilon^b_{p-1}[n-1]+ \gamma_{p-1} e^b_{p-1}[n-1] e^b_{p-1}[n-1]\,$\\
$e^f_{p} [n]= e^f_{p-1}[n]+ k^f_{p}[n-1] e^b_{p-1}[n-1]\, $\\
$e^b_{p}[n] = e^b_{p-1}[n-1]+  k^b_{p}[n-1] e^f_{p-1}[n]\, $\\      
$k^f_{p}[n] = k^f_p[n-1]-\frac{\gamma_{p-1}e^b_{p-1}[n]e^f_{p}[n]}{\epsilon^b_{p-1}[n]}\,$ \\
$k^b_p[l] = k^b_p[n-1]-\frac{\gamma_{p-1}e^f_{p-1}[n]e^b_{p}[n]}{\epsilon^f_{p-1}[n]}\, $\\
$\gamma_{p}=\gamma_{p-1}-\frac{\gamma_{p-1}^2|e^b_{p-1}[n-1]|^2}{\epsilon^b_{p-1}[n]}\, $ \\
   end\\
    $\sigma[l]=\sqrt{\epsilon^f_P[l]}/P$ \\
    $y[l]=e^f_P[l]$\\

Normalization for output\\
  $y[l]=y[l]/\sigma[l] $\\  
end\\

Initializations at $l=0$:\\
for $p=1,2,... P$\\
    $e^b_p[0] = 0$  $k^f_p[0] = 0$  $k^b_p[0] = 0$  $\gamma_p     = 1.0$ \\
    $\epsilon^f_p[0] = \delta$     $\epsilon^b_p[0] = \delta$ \\
being $\delta$ a value close to the average amplitude of the process.\\
\hline
\end{tabular}
\end{table}

 In the above relations no one of the parameters gives a direct estimation of the $\sigma$ of the guiding white noise process for the AR model.
We have to estimate it by using the quantities $\epsilon^f$ or $\epsilon^b$. In particular if we suspect to have non stationary data in which the overall RMS of the process changes in time we have to estimate it step by step.
The relation we used to estimate $\sigma$ in LSL filter is:
\begin{equation}
 \sigma[n]=\sqrt{\epsilon^f_P[n]}/P\, ,
\end{equation}
being $P$ the order we choose for the filter and consequently for the AR fit to the PSD.

Moreover we have to normalize this quantity with respect to the number of iterations we used to estimated it. If $\lambda=1$ we used all the data to achieve the converging value for $\sigma[n]$, so we have to divide this value at the step $N$, by the length $N$ of the time series.
If we used $\lambda <1$ we have to normalize by the window of data we used that is equal to $
  \frac{1}{1-\lambda}$, called the memory of the filter.

If the $\sigma$ varies in time we well find a $\sigma$ that follows the changes,  choosing a good value for the parameter $\lambda$.

The novelty in our algorithm is the introduction of a normalization of the output of the whitening filter in such a way to make the process white and stationary. We accomplish this task by estimating the $\sigma[n]$ at each step $n$ and the by diving the output $ e^f[n]$, which is our whitened time series, by $\sigma[n]$.

If we do not normalize the output of the whitening filter by the varying $\sigma$ we will have white non stationary data, but if we divide the output by $\sigma$ we will find white stationary data, that are what we need in applying optimal signal search filter for Gaussian and stationary data.
In the next section we show the results of the application of this modified version of LSL algorithms to simulated non stationary data.
\section{LSL: application to  non stationary noise data}
\label{sec:toy}
\subsection{Toy model I: varying the parameter}
We build an AR process of order $P=2$ simulating a power spectral density in which one resonance is present and we let varying the frequency of the peak changing in time the value of one of the two parameter. 
\begin{figure}[hb]
\begin{center}
\epsfig{file=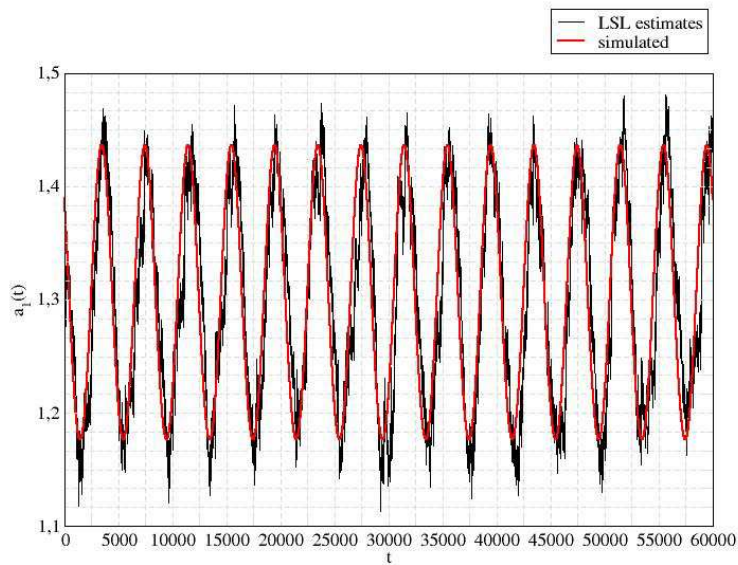,width=0.8\textwidth}
\end{center}
\caption{Simulated $a1(t)$ and LSL estimate}
\label{fig:ar1}
\end{figure}
\begin{figure}
\begin{center}
\epsfig{file=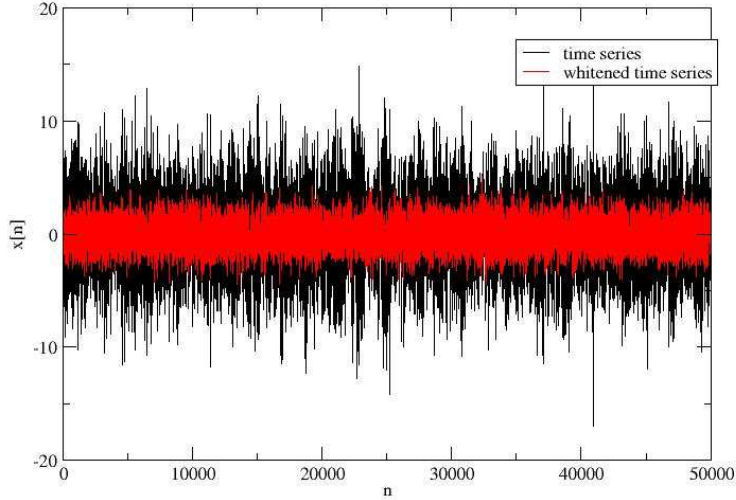,width=0.8\textwidth}
\end{center}
\caption{Time series $x[n]$ and LSL whitened one}
\label{fig:ar2}
\end{figure}
The value we use are the following:
\begin{equation}
A(1)= 1.3 \,\, A(2)=-0.9 \,\,\sigma=1.0 \,
\end{equation}
choosing a sampling frequency of $200\Hz$.

Moreover we let $A(1)$ vary in time with the following law:
\begin{equation}
A(1,t)=A(1) \exp(a\sin(2\pi\omega t +\phi)) \, ,  
\end{equation}
with the values $\phi=0$ $\omega=0.1 \Hz$ and $a=0.2$.

We  fit this process using LSL filter and whiten the data, using $\lambda=0.99$. In figure~\ref{fig:ar1} we show the simulated time varying parameter and the estimated one. In figure~\ref{fig:ar2} we show the simulated time series and the output of the LSL whitening filter.
\begin{figure}
\begin{center}
\epsfig{file=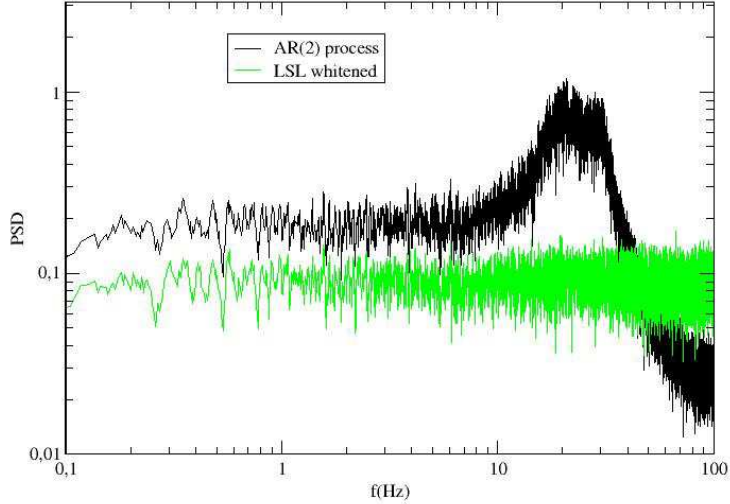,width=0.8\textwidth}
\end{center}
\caption{Power spectral density of the process $x[n]$ and LSL whitened one}
\label{fig:ar3}
\end{figure}

In these data it is one of the parameters of the AR model which changes its value in time, so when we estimate the reflection coefficients from the data we find also this variation in time if we use this forgetting factor $<1$. 
So the division of the output of the process by the  estimated $\sigma$ doesn't influence the whitening of the data, since the values of the $\sigma$ is constant in time.  

We check these results also by plotting the PSD at the input and at the output of the modified LSL filter and we plot the in figure~\ref{fig:ar3}. 
In this figure the peak of the PSD results broadened due to the moving of the main resonance,  while after the application of the LSL algorithm with forgetting factor $<1$ the PSD becomes flat, and also the non stationarity disappears.

\subsection{Toy model II:varying the $\sigma$}
We simulated an AR noise process in which the $\sigma$ of the guiding white normal noise changes in time with  the following function:
\begin{equation}
\sigma(t)=\sigma(1.0+(a\sin(2\pi\omega t)) \, ,  
\end{equation}
We use an AR(2) model with the same initial values of previous toy model,and the following values for the modulation
\begin{equation}
a= 0.4 \, \, \omega=0.005Hz  \, .
\end{equation}

\begin{figure}
\begin{center}
\epsfig{file=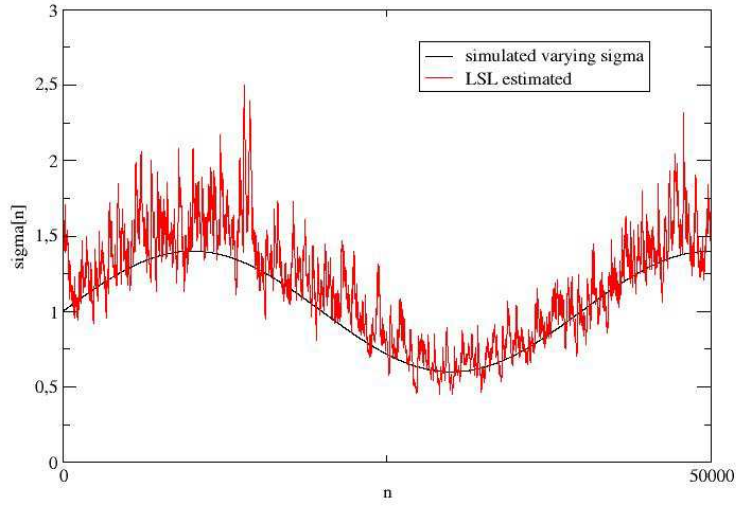,width=0.8\textwidth}
\end{center}
\caption{Simulated time varying $\sigma[n]$ and LSL estimated $\sigma$ with $\lambda=0.99$}
\label{fig:sigma}
\end{figure}

In this case it is crucial the division by the estimated $\sigma$ of the process if we want to have at the output of whitening filter stationary data.
In fact if we plot the estimated $\sigma$ of the LSL filter with $\lambda<1$, we find that, even if in a noisy way, the estimation follows the variations in time of the simulated $\sigma$ (see figure~\ref{fig:sigma}).

\begin{figure}
\begin{center}
\epsfig{file=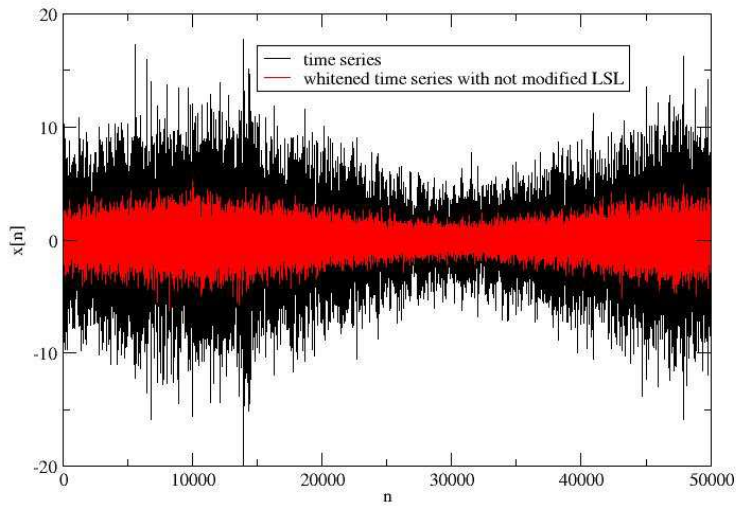,width=0.8\textwidth}
\end{center}
\caption{Simulated time series and not modified LSL whitening output}
\label{fig:sigma1}
\end{figure}

If we plot the output  of the standard implementation of the LSL whitening filter in figure~\ref{fig:sigma1}, it is evident the this filter has reduced the total RMS of the data, but it has not removed the modulation of the sigma of the process.

\begin{figure}
\begin{center}
\epsfig{file=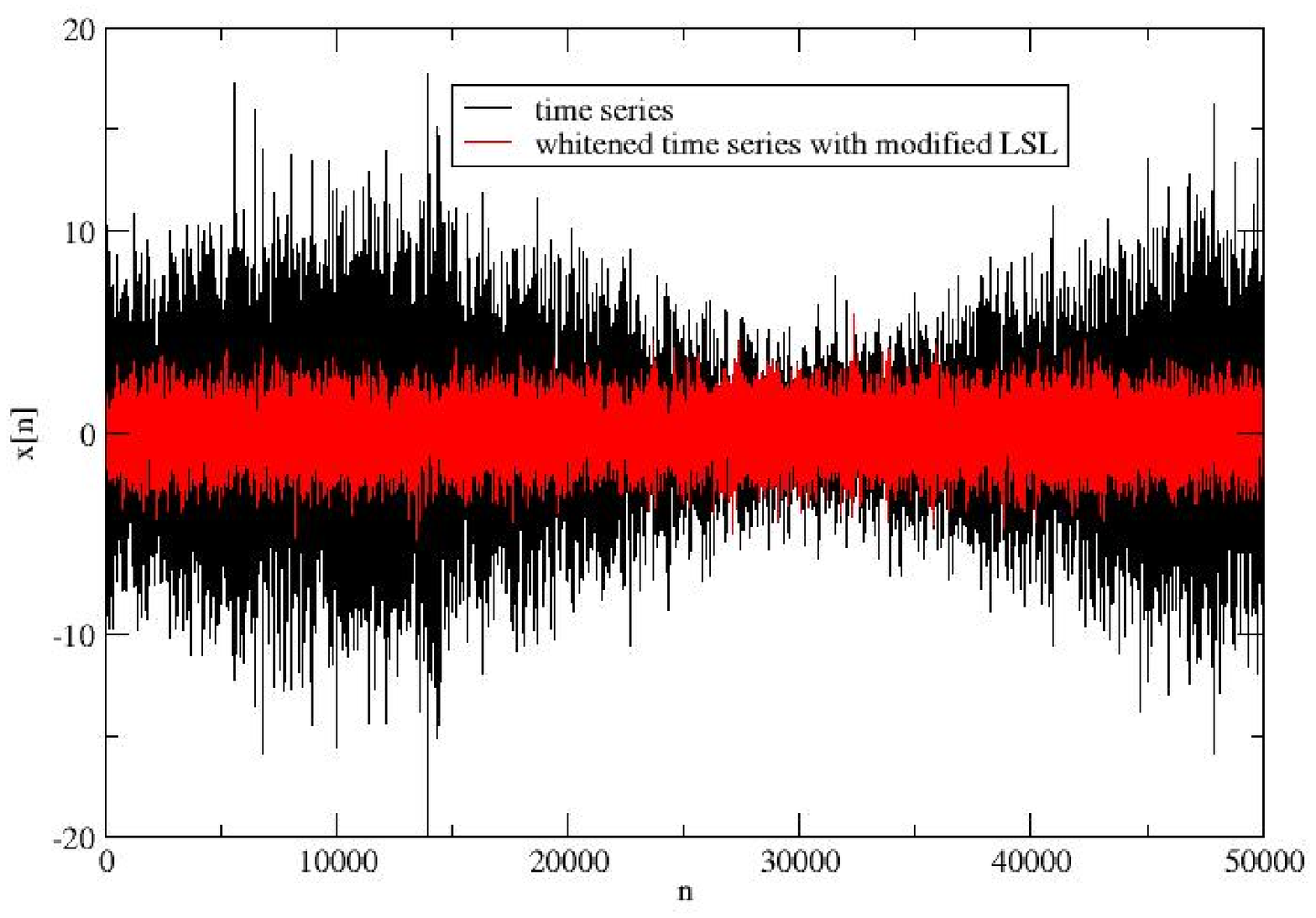,width=0.8\textwidth}
\end{center}
\caption{Time series $x[n]$ and modified LSL whitened one}
\label{fig:sigma2}
\end{figure}

If we apply the modified LSL filter, as it evident in figure~\ref{fig:sigma2}, we succeed in removing also the modulation of the data and in having a stationary whitened  time series.

\begin{figure}
\begin{center}
\epsfig{file=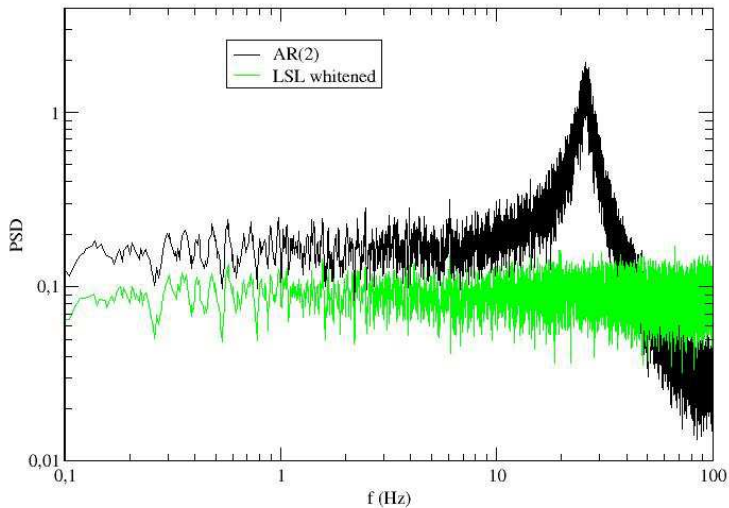,width=0.8\textwidth}
\end{center}
\caption{PSD of simulated non stationary AR(2) and modified LSL whitened one}
\label{fig:sigma3}
\end{figure}
As it is clear in figure~\ref{fig:sigma3} the whitening obtained with the modified LSL algorithm is good and the whitened PSD for non stationary data results flat.

\section{A realistic case: whitening the voice}
\label{sec:voice}

\begin{figure}
\begin{center}
\epsfig{file=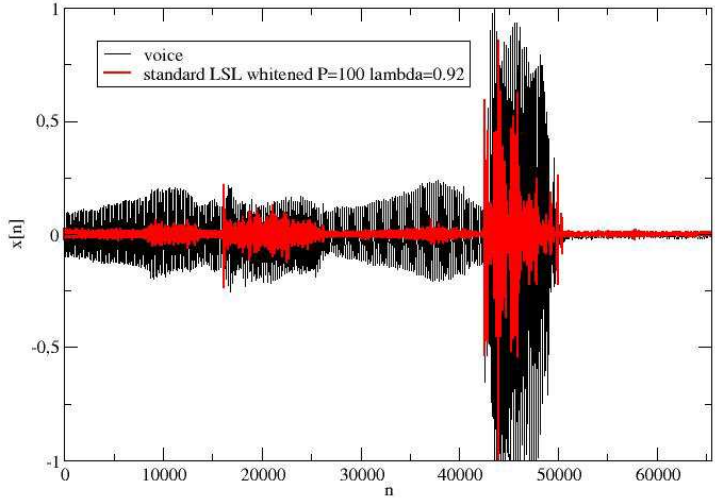,width=0.8\textwidth}
\end{center}
\caption{Voice time series in time domain  and  whitened ones.  The time series  was fitted as an AR(100) model and we use  forgetting factor $\lambda=0.92$ in the LSL algorithm}
\label{fig:voice1}
\end{figure}

\begin{figure}
\begin{center}
\epsfig{file=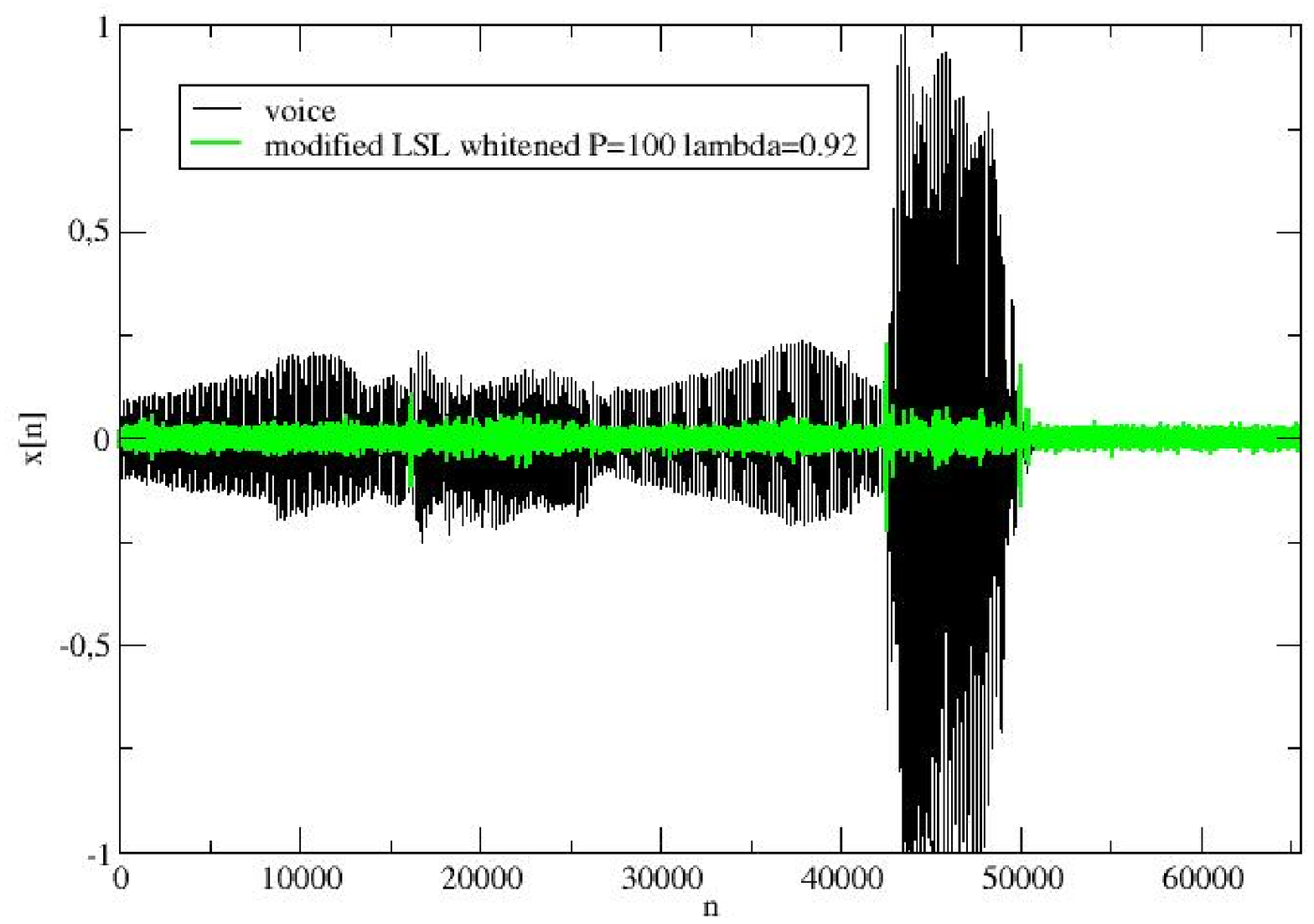,width=0.8\textwidth}
\end{center}
\caption{ Voice time series in time domain  and  whitened ones with modified algorithm.  The time series  was fitted as an AR(100) model and we use  forgetting factor $\lambda=0.92$ in the LSL algorithm}
\label{fig:voice4}
\end{figure}

\begin{figure}
\begin{center}
\epsfig{file=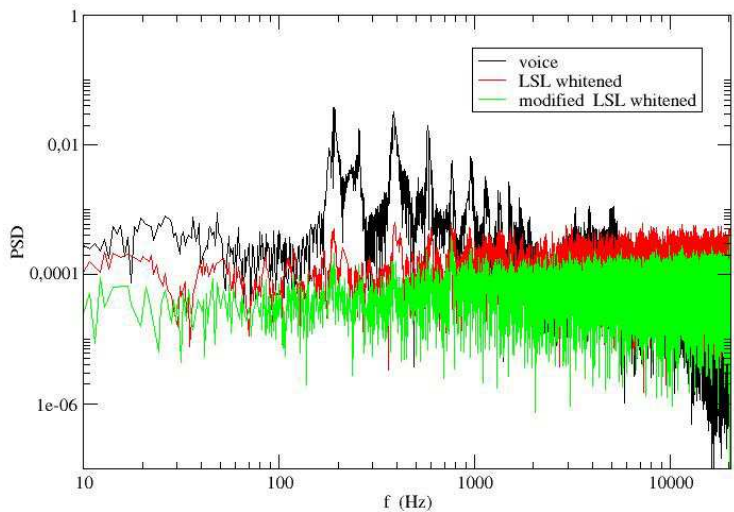,width=0.8\textwidth}
\end{center}
\caption{PSD for recorded voice, standard and modified LSL whitened ones with order $P=100$ and forgetting factor $\lambda=0.92$.}
\label{fig:voice2}
\end{figure}

In order to see the application of this algorithm on real data we perform a test on a recorded speech sound, that we convert in a time series.
 This is surely a non stationary time series, and we want to test if our algorithm is able to identify the speech and to remove all the features present in it.
This will also mean that we can be able to reconstruct the speech from the learned parameters~\cite{cuoco3}.
In figure~\ref{fig:voice4} we report the voice time series in time domain and the outputs of the standard and  LSL algorithm, using an order $P=100$ for the filter and a value $\lambda=0.92$ for the forgetting factor. 
If we do not apply the modification of the LSL filter, we succeeded in whitening the PSD (see figure~\ref{fig:voice2}), but non in removing the non stationarity, as is clear in figure~\ref{fig:voice1}. In figure~\ref{fig:voice4} we reports the results of the modified algorithm. The whitening results good and the variation of the RMS in time has disappeared. 

\begin{figure}
\begin{center}
\epsfig{file=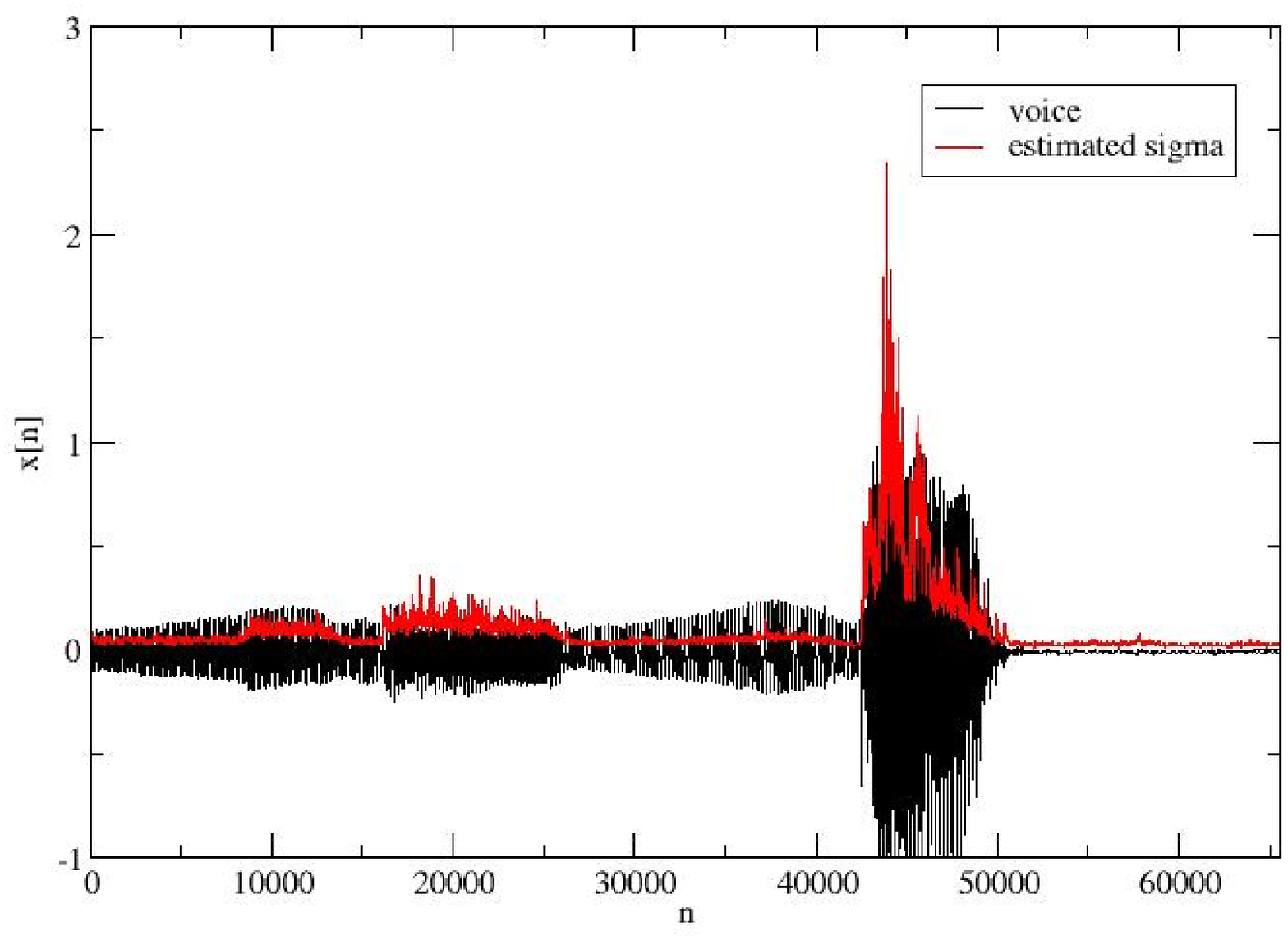,width=0.8\textwidth}
\end{center}
\caption{ Voice time series in time domain  and  estimated $\sigma[n]$.}
\label{fig:voice3}
\end{figure}

In figure~\ref{fig:voice3} we superimpose the estimated $\sigma[n]$ on the voice time series to show that most of the variation in time of the voice are due not to the variation of the AR parameters but to the variation of the $\sigma$ of the AR process.

These tests show that we have a powerful method to identify non stationary process and to whiten them. If we have the estimation of the parameters in time we can reconstruct step by step the original time series.

\section{Conclusion}
We build a whitening filter using a modified LSL algorithm to remove the non stationarity present in the RMS of the driving white noise for a simulated AR process. We find that with this algorithm is able to follow the non stationarity either coming from the AR parameters or from the $\sigma$ parameter and to remove them from the original data set. 

This kind of implementation could be useful if we want to deal with stationary and white data and we have to apply an optimal filter for signal detection.

Moreover we test this algorithm on speech signal, finding encouraging results on the identification of speech, so this application could be useful also in the reconstruction of speech phoneme.

\end{document}